# CX Lyrae 2008 Observing Campaign

**Pierre de Ponthière**
*15 Rue Pre Mathy, Lesve, Profondeville, 5170, Belgium*

**Jean-François Le Borgne**
*GEOS (Groupe Européen d'Observations Stellaires), 23 Parc de Levesville, 28300 Bailleau l'Evêque, France*

*and*

*Laboratoire d'Astrophysique de Toulouse-Tarbes, Observatoire Midi-Pyrénées (CNRS/UPS), Toulouse, France*

**F. -J. Hambsch**
*GEOS (Groupe Européen d'Observations Stellaires), 23 Parc de Levesville, 28300 Bailleau l'Evêque, France*

*Bundesdeutsche Arbeitsgemeinschaft für Veränderliche Sterne e.V. (BAV), Germany*

*and*

*Vereniging Voor Sterrenkunde, Werkgroep Veranderlijke Sterren (VVS), Belgium*

**Abstract**   The Blazhko effect in CX Lyr has been reported for the first time by Le Borgne *et al.* (2007). The authors have pointed out that the Blazhko period was not evaluated accurately due to dataset scarcity. The possible period values announced were 128 or 227 days. A newly conducted four-month observing campaign in 2008 (fifty-nine observation nights) has provided fourteen times of maximum. From a period analysis of measured times of maximum, a Blazhko period of 62 ± 2 days can be suggested. However, the present dataset is still not densely sampled enough to exclude that the measured period is still a modulation of the real Blazhko period. Indeed the shape of the (O–C) curve does not repeat itself exactly during the campaign duration.

## 1. Introduction

The star CX Lyr is classified in the *General Catalogue of Variable Stars* (GCVS; Samus *et al.* 2008), under number 520124, as an RRab variable star with a period of 0.61664495 day and with minimum and maximum magnitudes of 12.14 and 13.17V. CX Lyr is also identified as GSC 02121-02076 (STScI 2001). According to SIMBAD, CX Lyr is spectral type F4, and the (*B–V*) color index is thus around 0.4.

CX Lyr has been observed sporadically for over one hundred years. In total, fifty-one maxima have been recorded and data are available in the GEOS database (GEOS 2009). From these data, it can be seen that the pulsating period over the last century has decreased at a constant rate. Finally, an apparent Blazhko effect has been reported in CX Lyr for the first time by Le Borgne *et al.* (2007).

**2. New CCD observations and data reduction**

Hambsch and de Ponthière observed CX Lyr during fifty-nine nights between June 19, 2008 (JD 2454637), and November 12, 2008 (JD2454783). Hambsch's measurements were performed through *V* and $R_c$ photometric filters with an exposure time of 120 seconds. An ST8 CCD mounted on a C14 at *f*/5.6, located in Hechtel (Belgium), was used for imaging in 3 × 3 binning mode. Image acquisition was done using ccdsoft software(Software Bisque 2009). After dark and flat field correction the images were analysed with the muniwin software (Motl 2009). The positions and photometric data for the comparison and check stars used are listed in Table 1. The *V* magnitudes and color indexes were derived from the NOMAD catalogue (Zacharias *et al.* 2009). The color indexes have been obtained as differences between *B* and *V* values.

de Ponthière's measurements were performed through a *V* filter with an exposure time of thirty seconds using a Meade 8-inch LX200GPS at *f*/6.3 and ST7 CCD camera located in Lesve, Belgium. To improve the SNR, groups of four consecutive images were combined using the maximdl software (Diffraction Limited 2004). Aperture photometry was performed using custom software which evaluates the SNR and estimates magnitude errors in accordance with formulae (12) and (13) of Newberry (1991). The variable and comparison star data are listed in Table 2. The *V* magnitudes of C1 and C2 have been derived from simbad. The color indexes were obtained as differences between *B* and *V* values provided in the NOMAD catalogue.

The comparison star C1 was used as magnitude reference, C2 and C3 as check stars. To avoid measurements affected by variable sky conditions, the measurements with an estimated error greater than 0.050 magnitude (SNR < 21) have been discarded.

Hambsch and de Ponthière observed CX Lyr simultaneously on the night of September 1, 2009. From this common observation, it was possible to derive the magnitude offset created by the choice of different comparison stars. This offset has been used to adjust the magnitudes at maximum measured by Hambsch.

The list of measured maxima is provided in Table 3. This table includes an older measurement performed by Maintz (Hübscher, Steinbach, and Walter 2008). Unfortunately, the magnitude at maximum for this measurement cannot be reliably adjusted as a non-standard filter was used for this observation.

## 3. Period analysis and folded light curve

With the de Ponthière data, a period analysis performed with the anova algorithm of peranso (Vanmunster 2007) provided a pulsation period of $0.616703 \pm 0.000026$ day. The corresponding periodogram is presented in Figure 1a, the small peak at 0.7623 day (1.3107c/d) is a sampling alias. Aliases arise because the interval between measurements and signal period are of the same order. The alias frequency is related to signal one as $(f+1) / 2 = (1.6214+1) / 2 = 1.3107$, where $f = 1 / 0.616703$.

The folded light curves of de Ponthière's measurements are presented in Figure 2, with the elements HJD $2454677.5688 + 0.616703$ E (i.e., time of first maximum of de Ponthière). This figure clearly shows the existence of a Blazhko effect for this star.

A period analysis with the lomb-scargle algorithm in peranso has revealed side lobes close to the main lobe as shown in Figure 1b. The frequencies of two closer side lobes are, respectively: $f1 = 1.606875$ and $f2 = 1.635417$ day$^{-1}$. An estimation of the Blazhko period can be derived as $2 / (f2 - f1) = 70$ days. The anova algorithm reveals also equivalent side lobes but their levels are very weak.

A careful analysis of folded light curves shows that when the maximum is reached with maximal delay in respect to the ephemeris, the magnitude at the preceding minimum is at its lowest value. It was the case for JD 2454758, with a minimum magnitude of 13.37 measured at phase 0.84.

For de Ponthière's measurements, Le Borgne evaluated the times of maxima with a custom algorithm, fitting the measurement curve by a polynomial function whose degree depends on the number and dispersion of points. Hambsch evaluated the maxima from his own observations with the peranso program.

A linear regression on all (O–C) values presented in Table 3 provided a new pulsation period of $0.61675 \pm 0.000024$ day. The (O–C) values were re-evaluated with this new pulsation period. The elements we use for 2008 observations is then:

$$\text{HJD} = 2454677.5688 + 0.61675\text{E} \qquad (1)$$
$$\pm 0.0037 \ \pm 0.000024\text{E}$$

The Blazhko period was evaluated by an analysis of the (O–C) value variations with the anova algorithm. The resulting period is $62 \pm 2$ days. This value is probably more accurate than the value derived previously from the global spectral analysis. Indeed, the spectral analysis did not include the Hambsch observations. For those observations only the times of maxima and their corresponding magnitudes were available. However, it is too soon to announce an accurate Blazhko period for this star. The data cover only 2.5 Blazhko periods with a limited number of observations. It is still possible that the period detected here is a variation superimposed on the previously reported potential Blazhko periods of 227 and 128 days (Le Borgne *et al.* 2007), although the time variations of (O–C) and magnitude at maximum shown in Figure 3 leave little doubt.

The (O–C) values as a function of time are presented in Figure 3. The (O–C) values and magnitudes at maximum as a function of Blazhko phase are presented in Figure 4. The Blazhko phases are calculated with the elements 2454677.5688 + 62.0 E.

The curve of the (O–C) shows a slow change for increasing (O–C) values and an abrupt change for decreasing values. We can see that the periodic variations of (O–C) and magnitude are obviously synchronous and more or less in phase. The (O–C) values and magnitudes reach their minimum at the same time. But the magnitude maximum is reached while the (O–C) values are still increasing.

The Blazhko period can be divided into three sub-periods of more or less equal durations:

- (O–C) value and magnitude increase.
- (O–C) value increases and magnitude decreases.
- (O–C) value decreases abruptly and magnitude decreases.

The behavior for UX Tri as reported by Achtenberg and Husar (2006), presented in Figure 5, is opposite to the present findings. An abrupt variation for increasing (O–C) values and a slow change for decreasing values is the usual behavior of Blazhko (O–C) phase diagrams. The magnitude and (O–C) minima occur at the same phase, but the behavior for three sub-periods (of unequal duration) is as follows

- (O–C) value increases abruptly and magnitude increases.
- (O–C) value decreases and magnitude increases.
- (O–C) value decreases and magnitude decreases.

It is clear that although both CX Lyr and UX Tri have a strong Blazhko effect, they have different characteristics. Both variations of magnitude at maximum are close to sinusoidal and (O–C) are reversed. The behavior of UX Tri is typical of the Blazhko effect.

Further observations are required to be sure that the measured periodic variations are not a simple oscillation superimposed on the main Blazhko variation. With new observations, it will be possible to remove this doubt and obtain a better value of the Blazhko period.

## 4. Acknowledgements

We would like to thank Jacqueline Vandenbroere (GEOS) for her suggestions at the beginning of this CX Lyr campaign. Thanks to Dieter Husar for authorizing publication of the UX Tri figures. This work has made use of the SIMBAD database, operated at CSD, Strasbourg, Strasbourg, France, and the GEOS RR Lyr database.

Table 1. Comparison star data for CX Lyr (F. -J. Hambsch observations).

| Identification | R.A. (2000) h m s | Dec. (2000) ° ′ ″ | V | B–V |
|---|---|---|---|---|
| C1 GSC 2121-1980 | 18 51 14.24 | 28 43 37.88 | 12.74 | 0.55 |
| C2 GSC 2121-2842 | 18 51 07.00 | 28 45 12.88 | 13.0 | 0.70 |

Table 2. Comparison star data for CX Lyr (P. de Ponthière observations).

| Identification | R.A. (2000) h m s | Dec. (2000) ° ′ ″ | V | B–V |
|---|---|---|---|---|
| C1 GSC 2121-2818 | 18 51 51.49 | 28 49 08.11 | 10.58 | 0.58 |
| C2 GSC 2121-2053 | 18 51 16.38 | 28 51 16.38 | 10.48 | 0.49 |
| C3 GSC 2121-1980 | 18 51 14.24 | 28 43 37.88 | 12.74 | 0.55 |

Table 3. List of measured maxima of CX Lyr.

| Maximum HJD | Error | O–C (day) | E | Observer |
|---|---|---|---|---|
| 2452362.4056 | 0.0004 | –0.0019 | –511 | G. Maintz |
| 2454637.4929 | 0.0026 | 0.0138 | –65 | F. -J. Hambsch |
| 2454661.5051 | 0.0020 | –0.0278 | –26 | F. -J. Hambsch |
| 2454677.5688 | 0.0018 | 0.0000 | 0 | P. de Ponthière |
| 2454685.5930 | 0.0020 | 0.0063 | 13 | P. de Ponthière |
| 2454692.3815 | 0.0013 | 0.0104 | 24 | P. de Ponthière |
| 2454708.4197 | 0.0019 | 0.0127 | 50 | P. de Ponthière |
| 2454711.5050 | 0.0040 | 0.0142 | 55 | P. de Ponthière |
| 2454719.5020 | 0.0050 | –0.0068 | 68 | P. de Ponthière |
| 2454724.4300 | 0.0040 | –0.0129 | 76 | P. de Ponthière |
| 2454729.3630 | 0.0030 | –0.0140 | 84 | P. de Ponthière |
| 2454750.3518 | 0.0016 | 0.0048 | 118 | P. de Ponthière |
| 2454758.3736 | 0.0012 | 0.0086 | 131 | P. de Ponthière |
| 2454774.4100 | 0.0030 | 0.0092 | 157 | P. de Ponthière |
| 2454782.3940 | 0.0030 | –0.0248 | 170 | P. de Ponthière |

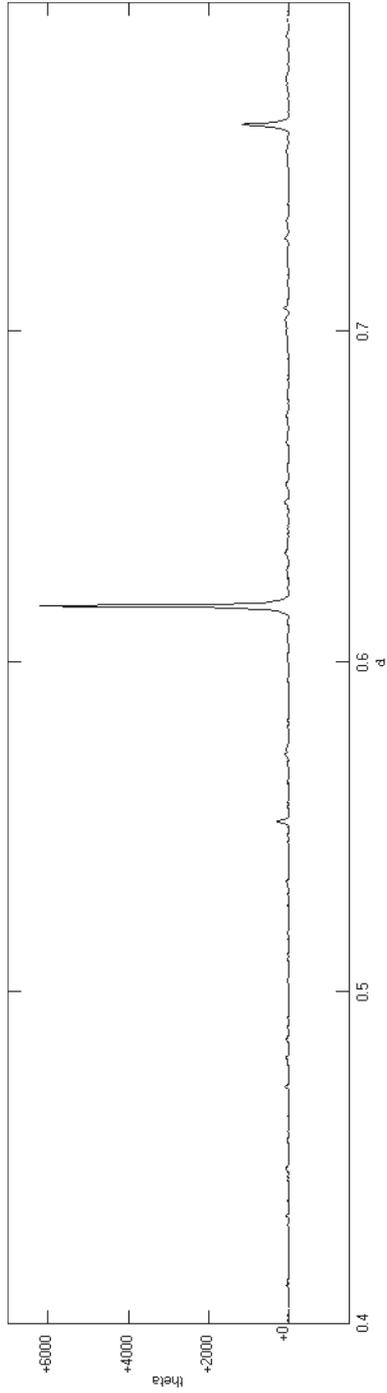

Figure 1a. CX Lyr Periodogram (PERANSO—ANOVA algorithm).

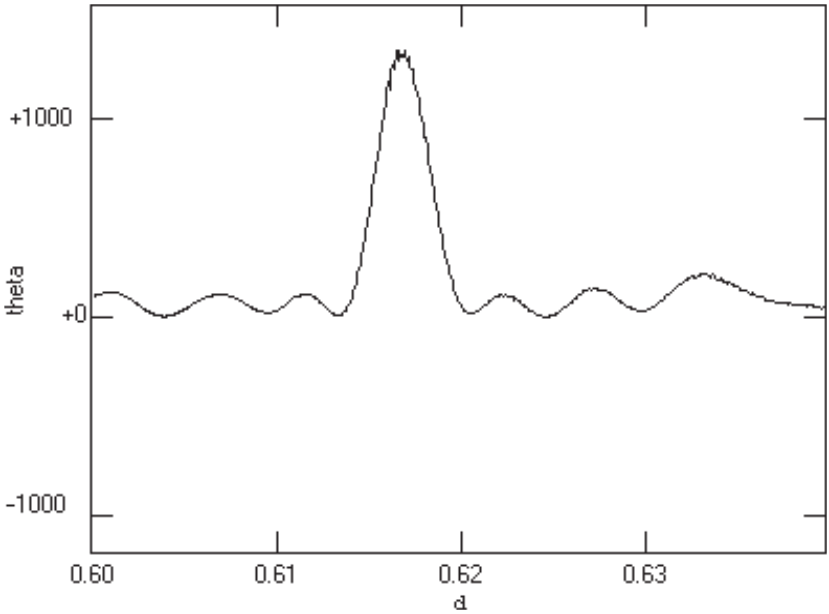

Figure 1b. CX Lyr Periodogram (PERANSO—LOMB-SCARGLE algorithm).

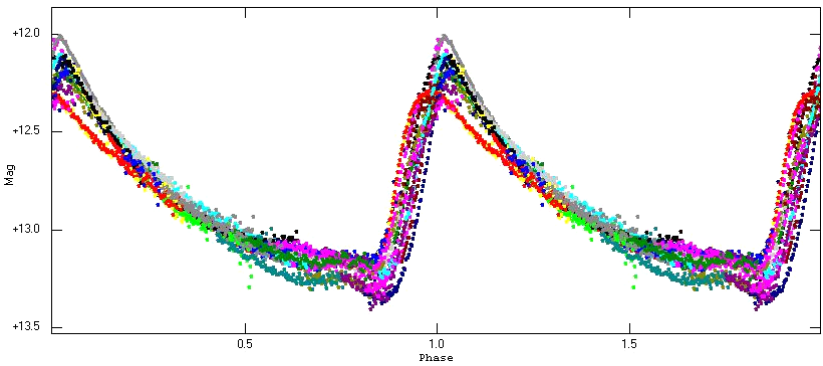

Figure 2. CX Lyr folded light curves (de Ponthière data).

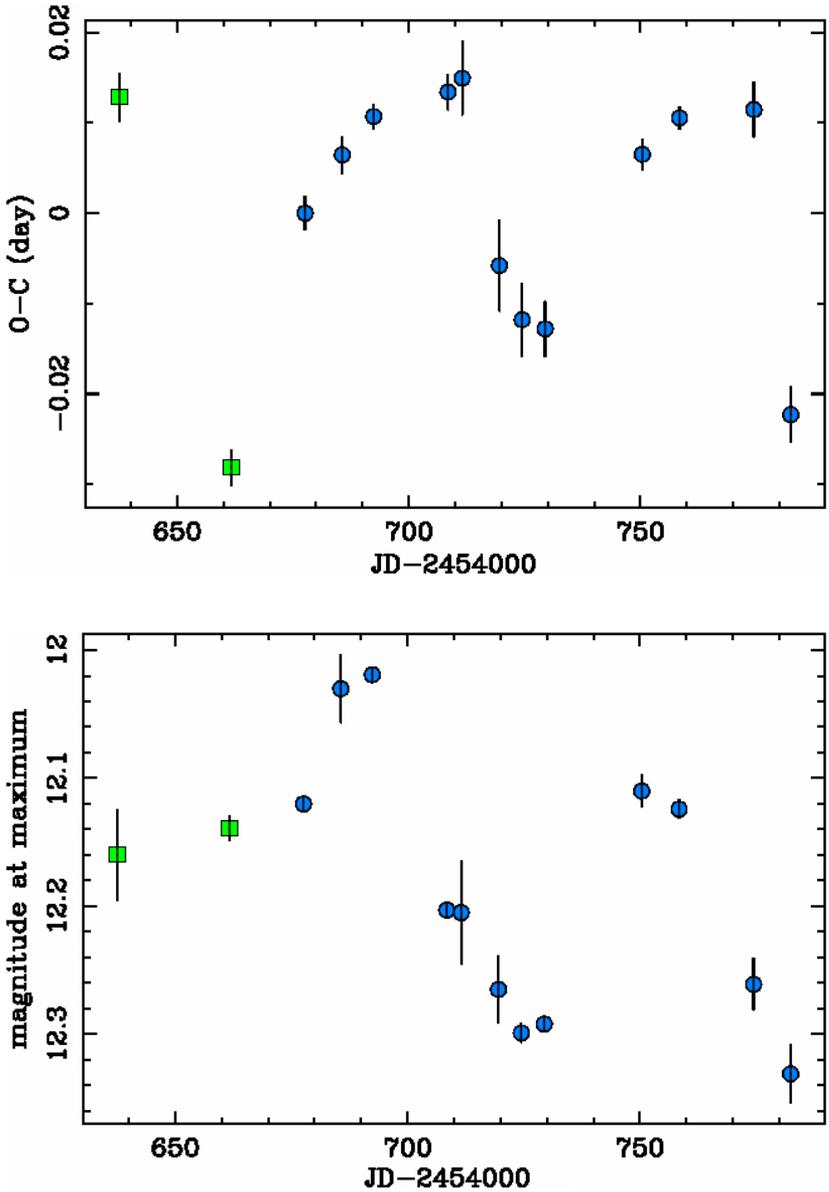

Figure 3a, 3b. CX Lyr (O–C) (top) and magnitude (bottom) at maximum diagram. Symbols correspond to observers: F.-J. Hambsch (squares) and P. de Ponthière (circles).

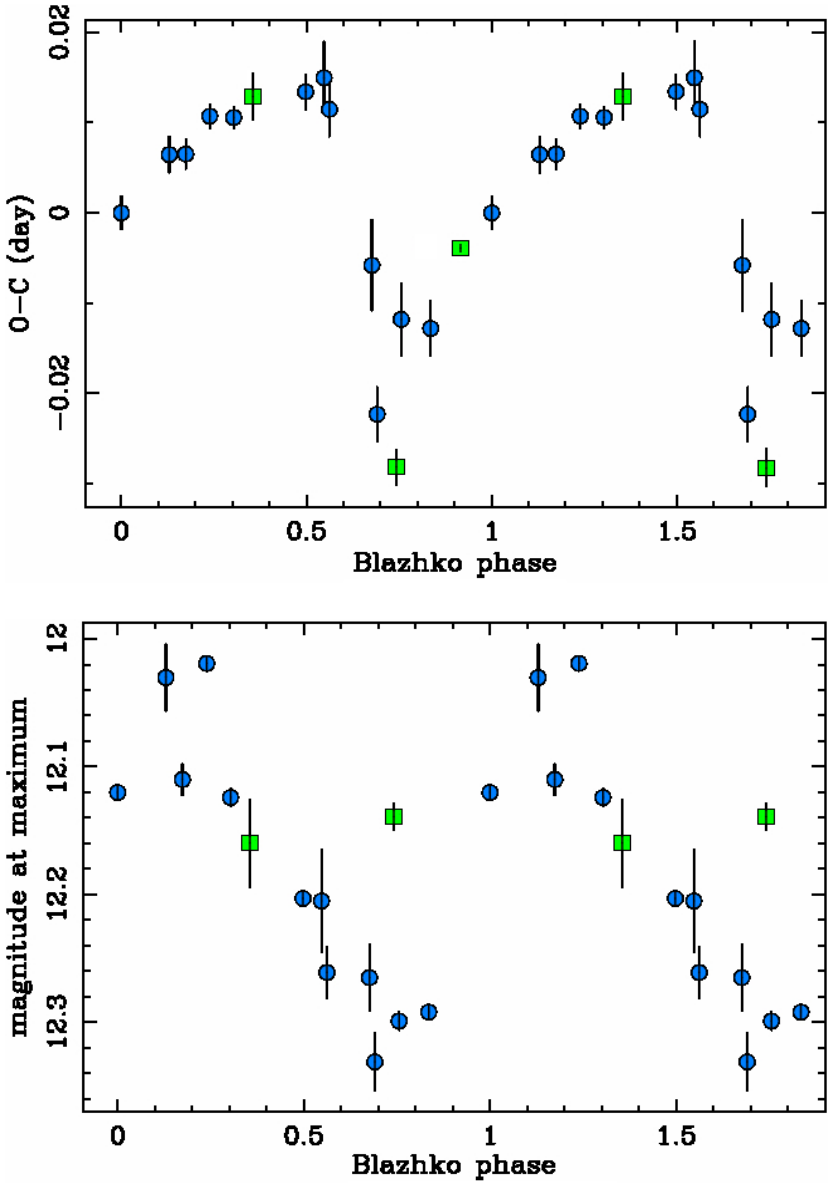

Figure 4a, 4b. CX Lyr (O–C) (top) and magnitude (bottom) at maximum versus Blazhko phase. Symbols correspond to observers: F.-J. Hambsch (squares) and P. de Ponthière (circles).

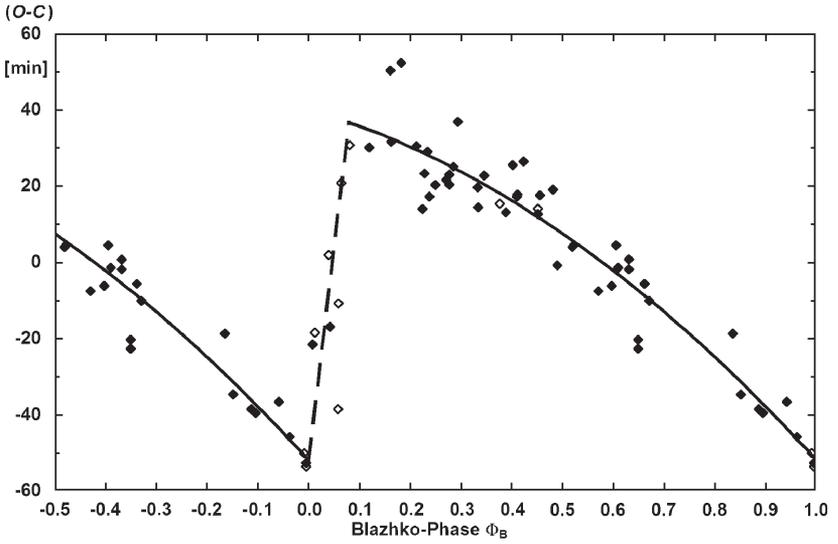

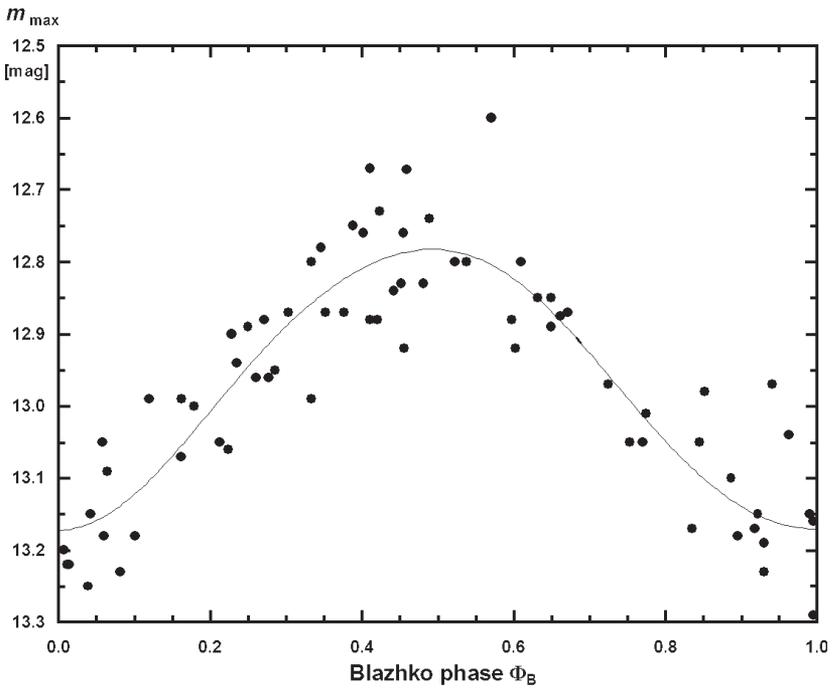

Figure 5a, 5b. UX Tri (O–C) (top) and magnitude (bottom) at maximum versus Blazhko phase.